\documentclass[prd,onecolumn]{revtex4}
\begin{document}
\title{Wormholes supported by phantom energy from Shan-Chen cosmological fluids}
\author{Deng Wang}
\email{Cstar@mail.nankai.edu.cn}
\affiliation{Theoretical Physics Division, Chern Institute of Mathematics, Nankai University,
Tianjin 300071, China}
\author{Xin-He Meng}
\email{xhm@nankai.edu.cn}
\affiliation{{Department of Physics, Nankai University, Tianjin 300071, P.R.China}\\{State key Lab of Theoretical Physics, Institute of Theoretical Physics, CAS, Beijing 100080, P.R.China}}

\begin{abstract}
The firm observational confirmation of the late-time acceleration of the universe expansion has proposed a major challenge to the theoretical foundations of cosmology and the explanation of the acceleration mechanism requires the introduction of either a simply cosmological constant, or of a mysterious dark energy component (time dependent or modified gravities), filling the universe and dominating its current expansionary evolution. Universally given that the universe is permeated by a dark energy fluid, therefore, we should also investigate the astrophysical scale properties from the dark energy effects. In the present paper, the exact solutions of spherically-symmetrical Einstein field equations describing wormholes supported by phantom energy that violates the null energy condition  from Shan-Chen fluid description are obtained. We have considered the important case that the model parameter $\psi\approx1$ which corresponds to the `` saturation effect ", and this regime corresponds to an effective form of `` asymptotic freedom " for the fluids, but occurring at cosmological rather than subnuclear scales. Then we investigate the allowed range values of the model parameters $g$ and $\omega$ when the space-time metrics describe wormholes and discuss the possible singularities of the solutions, finding that the obtained spacetimes are  geodesically complete. Moreover, we construct two traversable wormholes through matching our obtained  interior solutions to the exterior Schwarzschild solutions and calculate out the total mass of the wormhole when the wormhole throat size $r\leq a$ or $r\leq b$, respectively. Finally, we acquire that the surface stress-energy $\sigma$ is zero and the surface tangential pressure $\wp$ is positive when discussing the surface stresses of the solutions and analyze the traversable wormholes.
\end{abstract}
\maketitle
\section{Introduction}
Nowadays, all existing observational data are in good agreement with the simplest picture of six parameter base cosmology, namely, the $\Lambda$CDM model, favorably left a viable dark energy component to enrich a cosmological constant effects. According to this model, the universe is described well by a Friedmann-Robertson-Walker (FRW) metric, whose gravity source is a mixture of non-interacting perfect fluids including a cosmological constant with the main dark matter component. However, no theoretical model able to precisely  determine the nature of dark energy is available at present. In recent years, besides modified gravities there are a variety of models for dark energy proposed which include, for example, the quintessence model \cite{Turner,Caldwell}, scalar field models with nonstandard kinetic terms (k-essence) \cite{C.Armendariz-Picon V.Mukhanov and P.J.Steinhardt 2000}, the Chaplygin gas \cite{M.C.Bento O.Bertlami and A.A.Sen 2002}, braneworld models \cite{L.}, the dark fluid models \cite{rm}  and cosmological models from scalar-tensor theories of gravity (see, e.g., Refs.\cite{V.Sahni and A.A.Starobinsky 2006, P.Ruiz-Lapuente2007} and references therein)... Unfortunately, these models are all described mainly by the similar behaviour equation of state. Nevertheless, in 1993, a new  modified equation of state (EoS) was first proposed by Shan and Chen (SC) in the context of lattice kinetic theory \cite{X.Shan and H.Chen1993} for describing complex fluid, with the primary intent that repulsion is replaced by a density-dependent attraction. Then ten years later, Donato Bini et al have studied the dark energy properties from cosmological fluids obeying the SC non-ideal equation of state \cite{Donato Bini et al2013}. The main idea is to postulate the cosmological fluids obey SC equation with `` asymptotic freedom " regimes , namely, ideal gas behaviors at both high and low density regimes, with a liquid-gas coexistence cycle . Through some numerical calculations, they found that in the cosmological Friedmann-Robertson-Walker (FRW ) framework a cosmic fluid obeying the SC equation of state naturally evolves towards a present-day universe state with a suitable dark-energy component, with no need of invoking any cosmological constant. Therefore, we are very interested in exploring the SC EoS effects further in the astrophysics scales if the dark energy phenomena is universal, especially the possible wormholes supported by this new cosmological fluid as the phantom dark energy has violated the null energy condition that is essential for wormhole formation. Another motivation is that two recent studies \cite{Rahaman,P} have demonstrated the possible existence of wormholes in both the outer regions of the galactic halo and in the central parts of the halo, respectively, based on NFW (Navarro-Frenk-White) density profile and the URC (Universal Rotation Curves) dark matter model simulation and fittings \cite{R,NFW}. Especially, the second result is an important compliment to the earlier results on theoretical discussions on possible wormhole properties, thereby inspiring common interests in exploring the possible existence of wormholes in most of the spiral galaxies with the universally distributed dark energy.

Wormholes are possibly amazing result in the Einstein's gravity theory of General Relativity (GR), since they can provide an alternative method for rapid interstellar travel with novelty properties, which may be defined briefly as tunnels in the spacetime topology connecting different universes or widely separated regions of our universe via a throat \cite{M.S.Moris and K.S.Thorne1988,M.Visser}. Recently, the increasing attention to such objects investigation  is due in some sense to the discovery that our universe is undergoing an accelerated expansion \cite{Riess A.G.1998, Perlmutter S.J.1999}, coined as the dark energy phenomena. Since the two cases (wormholes and accelerated expansion universe) are both in violation of a null energy condition explicitly (The null energy condition requires the stress-energy tensor $T_{\alpha\beta}k^\alpha k^\beta\geq0$ for the all null vectors), $p+\rho < 0$, and consequently other energy conditions. Therefore, an interesting and implying overlap has appeared between the two seemingly separated subjects. More precisely, the accelerated universe expansion can be described globally by the GR strictly derived Friedmann equation $\ddot{a}/a=-\frac{4\pi}{3}(\rho+3p)$, i.e., we require $\ddot{a}>0$. (Here we take units throughout $G=c=1$). It is obvious from the previous works \cite{mw, Pe} that we know  the cosmic acceleration is caused mathematically by a hypothetical negative pressure dark energy with the energy density $\rho>0$ and $p=\omega\rho$ with the EoS parameter $\omega<-\frac{1}{3}$. And one important situation of dark energy modelings is the phantom dark energy when the EoS parameter $\omega<-1 $,  for which the same violation of the null energy condition as for the wormhole case occurs. Thus, we can explore the wormholes supported by phantom energy through studying the situation for $\omega<-1$. Additionally, in the quintessence models the parameter range is $-1<\omega<-1/3$, and recent astrophysics observations mildly favor phantom dark energy scenario. Moreover, the case $\omega=-1$ corresponds to a cosmological constant, and $\omega=-2/3$ is extensively analyzed in the nice work \cite{P.F.Gonzales-Diaz 2002}.

A recurring problem in general relativity is always to find the exact solutions to the Einstein field equations. In this paper, provided that the universe is permeated by a dark energy fluid, therefore, besides the cosmic scales we should also investigate the astrophysical scale properties of dark energy. So we plan to explore the phantom energy wormholes from SC fluids with considering Shan-Chen's interesting and intriguing EoS work applying into the astrophysics scales.

The current work is organized as follows. In the next section, we review some properties of cosmic wormhole model and make a brief introduction for SC equation of state. Moreover, we derive the field equation of the important case for the parameter $\psi=1$ starting from a general line element and SC equation of state. In the sub-sections of Section 3, we make two special choices for the redshift function: $\Phi=C$ and $\Phi(r)=\frac{1}{2}\ln(r_1/r)$, respectively, and acquire the corresponding two solutions. We analyze the singularities and construct two traversable wormhole through matching the interior solutions to the exterior Schwarzschild solutions. Moreover, we calculate out the total mass of the wormhole when $r\leq a$ or $r\leq b$, and find that the surface stress-energy $\sigma$ is zero and the surface tangential pressure $\wp$ is positive when discussing the surface stresses of the solutions. Subsequently, we analyze the traversable properties of the wormholes. In the final section, we make some conclusions and discussions about our work and point out the possible work in the future.

\section{The problem background}
\begin{flushleft}
The metric of the normal wormhole system can be generally written as
\end{flushleft}
\begin{equation}
ds^2=-e^{2\Phi(r)}dt^2+e^{\alpha(r)}dr^2+r^2(d\theta^2+\sin^2\theta d\theta^2),
\end{equation}
where $r$ is the radial coordinate which runs in the range $r_0\leq r<\infty$, $0\leq\theta\leq\pi$ and $0\leq\phi\leq2\pi$ are the angular coordinate. $\Phi(r)$ is redshift function, for it is related to the gravitational redshift. The function $\alpha(r)$ has a vertical asymptote at the throat $r=r_0$:
\begin{equation}
\lim_{r\rightarrow (r_0)+}\alpha(r)=+\infty.
\end{equation}
Its relationship to the shape function $b(r)$:
\begin{equation}
e^{2\alpha(r)}=\frac{1}{1-\frac{b(r)}{r}}.
\end{equation}
It follows that
\begin{equation}
b(r)=r(1-e^{-2\alpha(r)}).
\end{equation}
To describe a wormhole, the metric should obey some conditions \cite{M.S.Moris and K.S.Thorne1988}. The metric coefficient $e^{2\Phi(r)}$ should be finite and non-vanishing in the vicinity of $r_0$, $b(r)$ should satisfy the following relations:
\begin{equation}
b(r_0)=r_0,
\end{equation}
\begin{equation}
b'(r_0)<1,
\end{equation}
\begin{equation}
b(r)<r, r>r_0.
\end{equation}
Using Einstein field equation, $G_{\alpha\beta}=8\pi T_{\alpha\beta}$, in an orthonormal reference frame, we have the following stress-energy scenario
\begin{equation}
G_{tt}=8\pi T_{tt}=8\pi\rho=\frac{2}{r}e^{-2\alpha(r)}\alpha'(r)+\frac{1}{r^2}(1-e^{-2\alpha(r)}),
\end{equation}
\begin{equation}
G_{rr}=8\pi T_{rr}=8\pi p=\frac{2}{r}e^{-2\alpha(r)}\Phi'(r)-\frac{1}{r^2}(1-e^{-2\alpha(r)}).
\end{equation}
Then use SC equation
\begin{equation}
p=\omega\rho_{(crit),0}[\frac{\rho}{\rho_{(crit),0}}+\frac{g}{2}\psi^2],
\end{equation}
\begin{equation}
\psi=1-e^{-\beta\frac{\rho}{\rho_{(crit),0}}}.
\end{equation}
Where $\rho_{(crit),0}=3(H_0)^2/8\pi$ is the present value of the critical density ($H_0$ denoting the Hubble constant) and the dimensionless quantities $\omega$, $g\leq0$ and $\beta\geq0$ can be regarded as free parameters of the model.

Substitution yields
\begin{equation}
\omega\alpha'(r)+\frac{\omega+1}{2r}(e^{2\alpha(r)}-1)+2\pi\omega\rho_{(crit),0}ge^{2\alpha(r)}[1-e^{-\beta\frac{\frac{2}{r}e^{-2\alpha(r)\alpha'(r)+\frac{1}{r^2}(1-e^{-2\alpha(r)})}}{8\pi\rho_{(crit),0)}}}]^2=\Phi'(r).
\end{equation}
We find two solutions when $\rho\gg\rho_\ast$, $\rho_\ast=\rho_{(crit),0)}/\beta$ ($\rho_\ast$ being the typical density above which $\psi$ undergoes a ``saturation effect"), i.e., $\psi\approx1$ \cite{Donato Bini et al2013}, this regime corresponds to an effective form of asymptotic freedom. So the SC equation of state (10) becomes
\begin{equation}
p=\omega\rho_{(crit),0}[\frac{\rho}{\rho_{(crit),0}}+\frac{g}{2}].
\end{equation}
Since the pressure $p$ is negative, so $-\rho<\frac{g}{2}\rho_{(crit),0}<0$, so $\omega<-\frac{\rho}{\rho+\frac{g}{2}\rho_{(crit),0}}<-1$ corresponds to a SC's version of phantom energy.

It follows that
\begin{equation}
\omega\alpha'(r)+\frac{\omega+1}{2r}(e^{2\alpha(r)}-1)+2\pi\omega\rho_{(crit),0}ge^{2\alpha(r)}=\Phi'(r).
\end{equation}
This equation tells us the close relationship between $\Phi'(r)$ and $\alpha'(r)$ and hence between $\Phi(r)$ and $\alpha(r)$.
It's easily to be seen when $\beta =0$ or $g=0$ the equation of state will reduce to $p=\omega\rho$.
\section{Exact solutions for wormholes}
\subsection{Special choice for the redshift function: $\Phi=C$}
We have the first solution through inserting redshift function by hand $\Phi'(r)=0$, resulting in $\Phi=constant=C$ \cite{Lobo F S N 2005},
then equation $(13)$ becomes
\begin{equation}
\omega\alpha'(r)+\frac{\omega+1}{2r}(e^{2\alpha(r)}-1)+2\pi\omega\rho_{(crit),0}ge^{2\alpha(r)}=0.
\end{equation}
Under initial condition $\alpha(r_0)=+\infty$. The solution is
\begin{equation}
e^{2\alpha(r)}=\frac{1}{r^{-1-\frac{1}{\omega}}[r^{1+\frac{1}{\omega}}-r_0^{1+\frac{1}{\omega}}+4\pi g\rho_{(crit),0}\frac{\omega(r^{2+\frac{1}{\omega}}-r_0^{2+\frac{1}{\omega}})}{1+2\omega}]}.
\end{equation}

So, the element is
\begin{equation}
ds^2=-e^{2C}dt^2+\frac{1}{r^{-1-\frac{1}{\omega}}[r^{1+\frac{1}{\omega}}-r_0^{1+\frac{1}{\omega}}+4\pi g\rho_{(crit),0}\frac{\omega(r^{2+\frac{1}{\omega}}-r_0^{2+\frac{1}{\omega}})}{1+2\omega}]}dr^2+r^2(d\theta^2+\sin^2\theta d\phi^2).
\end{equation}
Since $\Phi=C$, so $e^{2\Phi}\neq0$ and the sapcetime avoid an event horizon.
It's easily checked that $b(r_0)=r_0$, and if we want the shape function satisfies $b'(r_0)<1$ and $b(r)<r$, the parameters $g$ and $\omega$ must satisfy one relation
\begin{equation}
\frac{-1-\frac{1}{\omega}}{4\pi\rho_{(crit),0}r_0}<g\leq0.
\end{equation}
Eq.(18) is obtained by the simplest condition that the function
\begin{equation}
f(r)=r-b(r)=r^{-\frac{1}{\omega}}[r^{1+\frac{1}{\omega}}-r_0^{1+\frac{1}{\omega}}+4\pi g\rho_{(crit),0}\frac{\omega(r^{2+\frac{1}{\omega}}-r_0^{2+\frac{1}{\omega}})}{1+2\omega}].
\end{equation}
 $(f(r_0)=0)$ is monotonically increasing in the range $r>r_0$ and $b'(r_0)=-\frac{1}{\omega}-4\pi\rho_{(crit),0}<1$.
(We also observe that if $\omega>-1$ the flare-out condition is no longer satisfied).

Also clearly,the metric is not asymptotically flat. However, it can be glued to the external Schwarzschild solution
\begin{equation}
ds^2=-(1-\frac{2M}{r})dt^2+(1-\frac{2M}{r})dr^2+r^2(d\theta^2+\sin^2\theta d\theta^2).
\end{equation}
To match the interior to the exterior, one needs to apply the junction conditions that follow the theory of general relativity. If there no surface stress-energy  terms at the surface S, the junction is called a boundary surface. If, on the other hand, surface stress-energy terms are present, the junction is called a thin-shell.

A wormhole with finite dimensions, in which the matter distribution extends from the throat, $r=r_0$, to a finite distance $r=a$, obeys the condition that the metric is continuous. Due to the spherical symmetry the components $g_{\theta\theta}$ and $g_{\phi\phi}$ are already continuous,so one needs to impose continuity only on the remaining components at $r=a$:
\begin{equation}
g_{tt(int)}(a)=g_{tt(ext)}(a),
\end{equation}
\begin{equation}
g_{rr(int)}(a)=g_{rr(ext)}(a).
\end{equation}
Eqs.(18) and (19) are for the interior and exterior components, respectively. These requirements, in turn, lead to
\begin{equation}
\Phi_{int}(a)=\Phi_{ext}(a),
\end{equation}
\begin{equation}
b_{int}(a)=b_{ext}(a),
\end{equation}
Particularly,
\begin{equation}
e^{2\alpha(r)}=\frac{1}{1-\frac{b(a)}{a}}=\frac{1}{1-\frac{2M}{a}}.
\end{equation}

So, one can deduce the mass of the wormhole given by
\begin{equation}
M=\frac{1}{2}b(a)=\frac{r_0}{2}(\frac{r_0}{a})^{\frac{1}{\omega}}-2\pi g\rho_{(crit),0}\frac{\omega}{1+2\omega}[a^2-(r_0)^2(\frac{r_0}{a})^{1+\frac{1}{\omega}}].
\end{equation}
Return to $\Phi=C$,we now have
\begin{equation}
C'=e^{2\Phi}=1-(\frac{r_0}{a})^{1+\frac{1}{\omega}}+4\pi g\rho_{(crit),0}\frac{\omega}{1+2\omega}[a-r_0(\frac{r_0}{a})^{1+\frac{1}{\omega}}].
\end{equation}
Thus, the line element becomes
\begin{equation}
ds^2=-C'dt^2+\frac{1}{V(r)}dr^2+r^2(d\theta^2+\sin^2\theta d\phi^2),
\end{equation}
\begin{equation}
V(r)=r^{-1-\frac{1}{\omega}}[r^{1+\frac{1}{\omega}}-r_0^{1+\frac{1}{\omega}}+4\pi g\rho_{(crit),0}\frac{\omega(r^{2+\frac{1}{\omega}}-r_0^{2+\frac{1}{\omega}})}{1+2\omega}].
\end{equation}
Our next step is to take into account the surface stresses. Using the Daromis-Israel formalism \cite{N.Sen 1924, K.Lanczos 1924}, the surface stresses are given by
\begin{equation}
\sigma=-\frac{1}{4\pi a}(\sqrt{1-\frac{2M}{a}}-\sqrt{1-\frac{b(a)}{a}})
\end{equation}
and
\begin{equation}
\wp=\frac{1}{8\pi a}(\frac{1-\frac{M}{a}}{\sqrt{1-\frac{2M}{a}}}-[1+a\Phi'(a)]\sqrt{1-\frac{b(a)}{a}}).
\end{equation}

It is so clear that the surface stress-energy $\sigma$ is zero and the surface tangential pressure $\wp$ is positive.

Another important consideration is singularities of the solution. To see this, we can deduce that $V(r)=0$ when $r=r_0$, but it's not a physical singularity. Then let us consider a free particle with an energy $E=-u_0$ and angular momentum $L=u_{\phi}$ (in dimensionless units) moving in the Eq.(16), $u^{\mu}$ being the 4-velocity. As usual, we choose the plane to be $\theta=\pi/2$, so one can easily obtain from the condition $g_{\mu\nu}u^{\mu}u^{\nu}=\varepsilon$ ($\varepsilon=0$ for massless particles and $\varepsilon=-1$ for massive particles) that
\begin{equation}
(u^r)^2=V(r)(\frac{E^2}{-e^{2C}}-\frac{L^2}{r^2}+\varepsilon),
\end{equation}
Here we just consider radial null geodesics ($L=0, \varepsilon=0$), and let $E=1, \omega=-11/10, -e^{2C}=1, r_0=1, g=-7/1000, \rho_{(crit),0)}=1$ (Here we are just to demonstrate that the proper time is finite at the throat through taking these values which may be not true values for the parameters and quantities, for references, see \cite{Ravi, O.B.Zaslavskii 2005}). Hence, the above equation can be rewritten as
\begin{equation}
(\frac{dr}{d\tau})^2=\frac{-1+r^{\frac{1}{11}}-\frac{7\pi}{330}(-1+r^\frac{12}{11})}{r^{\frac{1}{11}}},
\end{equation}
$\tau$ is the proper time and solve this equation through some numerical calculations
\begin{equation}
r=InverseFunction[\int^m_1\frac{1}{330-\frac{330}{n^{\frac{1}{11}}}+\frac{7\pi}{n^{\frac{1}{11}}}-7\pi n}dn][\pm\frac{\tau}{\sqrt{330}}+D],
\end{equation}
where $m$ ,$n$ and $D$ is upper limit of integration, variable of integration and integration constant, respectively. Obviously, $\tau$ is finite when $r=r_0=1$ and $r=r_0$ is not physical singularity. Thus, one can easily find that the metric is geodesically complete.

After matching our interior solution to the exterior Schwarzschild solution, we have construct a traversable wormhole. So an important consideration affecting the traversability is the proper distance $l(r)$ from the throat to a point away from the throat:
\begin{equation}
l(r)=\int^r_{r_0}\frac{1}{\sqrt{V(r)}}dr,
\end{equation}
 which is finite through some simple numerical calculations.

Another consideration is time dilation near the throat. Here we let $v=dl/d\tau$ in order that $d\tau=dl/v$ (assuming that the observer who will pass through the wormhole in the spaceship has a non-relativistic speed, i.e.$\gamma=\sqrt{1-(v/c)^2}\approx1$). Because $dl=e^{\alpha(r)}$ and $d\tau=e^{\Phi(r)}$, we have for any coordinate interval $\triangle t$:
\begin{equation}
\triangle t=\int^{l_b}_{l_a}e^{-\Phi(r)}\frac{dl}{v}=\int^{r_b}_{r_a}\frac{1}{v}e^{-\Phi(r)}e^{\alpha(r)}dr.
\end{equation}
Going from the throat to $r$, we get
\begin{equation}
\triangle t=\int^r_{r_0}\frac{1}{v}\frac{1}{\sqrt{C'V(r)}}dr
\end{equation}
Through some numerical calculations in which we make some appropriate choices for parameters $\omega, r_0$ and $g$, we find that $\triangle t$ also behaves well near the throat.

\subsection{Special choice for the redshift function: $\Phi(r)=\frac{1}{2}\ln(r_1/r)$}
Another possibility is $\Phi(r)=\frac{1}{2}\ln(r_1/r)$ \cite{O.B.Zaslavskii 2005}, for some constant $r_1$, after substitution into Eq.(15) we have
\begin{equation}
\omega\alpha'(r)+\frac{\omega+1}{2r}(e^{2\alpha(r)}-1)+2\pi\omega\rho_{(crit),0}ge^{2\alpha(r)}=\frac{1}{2r}.
\end{equation}
Here we still consider the important case $\psi\approx1$, since the quantity $\psi$ can be interpreted as the density of a chameleon scalar field \cite{J.Khoury and A.Weltman 2004} which is used for reconciling large coupling models with local gravity constraints.
It follows that
\begin{equation}
e^{2\alpha(r)}=\frac{1}{r^{-1-\frac{2}{\omega}}[\frac{(1+\omega)(r^{1+\frac{2}{\omega}}-r_0^{1+\frac{2}{\omega}})}{2+\omega} +2\pi g\rho_{(crit),0}(2+\omega)\frac{\omega(r^{2+\frac{2}{\omega}}-r_0^{2+\frac{2}{\omega}})}{(1+\omega)(1+2\omega)}]}.
\end{equation}
So the line element becomes
\begin{equation}
ds^2=-\frac{r_1}{r}dt^2+\frac{1}{r^{-1-\frac{2}{\omega}}[\frac{(1+\omega)(r^{1+\frac{2}{\omega}}-r_0^{1+\frac{2}{\omega}})}{2+\omega} +2\pi g\rho_{(crit),0}(2+\omega)\frac{\omega(r^{2+\frac{2}{\omega}}-r_0^{2+\frac{2}{\omega}})}{(1+\omega)(1+2\omega)}]}dr^2+r^2(d\theta^2+\sin^2\theta d\theta^2).
\end{equation}
Here we have
\begin{equation}
b'(r_0)=-\frac{1}{\omega}-4\pi g\rho_{(crit),0}\frac{(2+\omega)r_0}{1+2\omega}<1,
\end{equation}
and assuming the simplest condition that
\begin{equation}
f(r)=r-b(r)=r-\frac{1+\omega}{2+\omega}r+\frac{1+\omega}{2+\omega}r^{-\frac{2}{\omega}-1}r_0^{\frac{2}{\omega}+1}-2\pi g\rho_{(crit),0}(2+\omega)\frac{2\omega r+2r^{-\frac{2}{\omega}-1}r_0^{\frac{2}{\omega}+2}}{(1+\omega)(1+2\omega)}
\end{equation}
is monotonically increasing in the range $r>r_0$ and $f(r_0)=0$.
Hence,we have the relation
\begin{equation}
\frac{-(1+\omega)(1+2\omega)}{4\pi\rho_{(crit),0}r_0\omega(2+\omega)}<g\leq0,     \omega<-2
\end{equation}
After easily checking, the metric satisfies the Eqs.(5-7) and is also not asymptotically flat. Besides, we have
\begin{equation}
r=InverseFunction[\int^m_1\frac{1}{\sqrt{2000n-2000n^{\frac{20}{11}}-2079\pi(n^{\frac{20}{11}}+n^2)}}dn][\pm\frac{\tau}{60\sqrt{5}}+D']
\end{equation}
where $m$ ,$n$ and $D'$ is upper limit of integration, variable of integration and integration constant, respectively. Similarly, here we still consider radial null geodesics ($L=0,\varepsilon=0$), and let $E=1, \omega=-11/10, -e^{2C}=1, r_0=1, g=-7/1000, \rho_{(crit),0)}=1$,
we find that the spacetime of the solution is also geodesically complete.

Once again, the metric Eq.(35) can be glued to the Schwarzschild region at $r=b$. Using Eq.(25), we have the mass of the wormhole
\begin{equation}
M=\frac{b+(1+\omega)(\frac{r_0}{b})^\frac{2}{\omega}r_0}{2(2+\omega)}-\pi g\rho_{(crit),0}\frac{(2+\omega)\omega[b^2-(\frac{r_0}{b})^\frac{2}{\omega}r_0^2]}{(1+\omega)(1+2\omega)}.
\end{equation}
Return to $\Phi(r)=\frac{1}{2}\ln(r_1/r)$, we have
\begin{equation}
r_1=b-2M=\frac{(1+\omega)[1-(\frac{r_0}{b})^\frac{2}{\omega}r_0]}{2+\omega}+2\pi g\rho_{(crit),0}\frac{(2+\omega)\omega[b^2-(\frac{r_0}{b})^\frac{2}{\omega}r_0^2]}{(1+\omega)(1+2\omega)}.
\end{equation}
Here let $C_1=r_1/r$,
so the line element becomes
\begin{equation}
ds^2=-C_1dt^2+\frac{1}{r^{-1-\frac{2}{\omega}}[\frac{(1+\omega)(r^{1+\frac{2}{\omega}}-r_0^{1+\frac{2}{\omega}})}{2+\omega} +2\pi g\rho_{(crit),0}(2+\omega)\frac{\omega(r^{2+\frac{2}{\omega}}-r_0^{2+\frac{2}{\omega}})}{(1+\omega)(1+2\omega)}]}dr^2+r^2(d\theta^2+\sin^2\theta d\theta^2).
\end{equation}
Then take into account the the surface stresses in the forms of Eqs.(30-31) again, the same conclusion that the surface stress-energy $\sigma$ is zero and the surface tangential pressure $\wp$ is positive is obtained. Similarly, doing the same to the metric as the first, we can also find that the wormhole is traversable and the proper distance $l(r)$ is finite and coordinate interval $\triangle t$ is well behaved near the throat.

\section{Conclusion and discussions}
To summarize, the explanation of the cosmic acceleration expansion requires the introduction of either cosmological constant, or of a mysterious component so called as dark energy, filling the universe and dominating its expansionary evolution currently. Given that the universe is permeated by a dark energy fluid smoothly, therefore, we should also  investigate the astrophysical scale properties of dark energy. In the first place, we have considered the important case $\psi\approx1$ in Shan-Chen cosmological modeling  which corresponds to the `` saturation effect '', and this regime corresponds to an effective form of  the `` asymptotic freedom " in the fluids, occurring at cosmological rather than subnuclear scales. Whence we find out two simple solutions of spherically-symmetrical Einstein equations with the SC equation of state describing a wormhole. Then, we explore the  value ranges of the parameters $g$ and $\omega$ when the space-time metrics describe wormholes and discuss the singularities of the solutions, and we find that the spacetimes of the two new solutions are both geodesically complete. In addition, we construct two traversable wormholes through matching our interior solutions to the exterior Schwarzschild solutions and calculate out the total mass of the wormhole when $r\leq a$ or $r\leq b$, respectively. Finally, we acquire that the surface stress-energy $\sigma$ is zero and the surface tangential pressure $\wp$ is positive when discussing the surface stresses of the solutions and analyze the traversability of the wormholes.

The century celebrating GR is still full of implying puzzlings, especially when confronting the dark matter and dark energy mysteries in various scales if the dark sectors are universal.
Wormholes are theoretically objects in Universe which now appear to attract more observational astrophysics interests and may provide a new window  for new physics.
The further work could be to explore the wormhole dynamics relations with the analogous studies of black holes by considering an obvious sensible relation between a transverse pressure and the energy density and to make some special choices for the shape function $b(r)$ as well. Moreover, it is worth to analyze the geodesic motion in the wormhole spacetime from SC cosmological fluids in details.   \\

\begin{center}
\textbf{Acknowledgments}
\end{center}

For helpful discussions and comments, we thank Prof. Saibal Ray and Jingling Chen, Qixiang Zou, Guang Yang, Shengsen Lu, and Liwei Yu. This work is partly supported by the National Science Foundation of China.

\end{document}